\documentclass[english,a4paper,pre,twocolumn,longbibliography,amsmath,amssymb]{revtex4-1}
\usepackage{graphicx,color,soul}

\newcommand{\dd}{\mathrm{d}}
\newcommand{\td}[2]{\frac{\dd #1}{\dd #2}}
\newcommand{\pd}[2]{\frac{\partial #1}{\partial #2}}
\newcommand{\mean}[1]{\langle #1 \rangle}
\newcommand{\Int}[1]{\int\dd #1\;}
\newcommand{\IInt}[3]{\int_{#2}^{#3}\dd #1\;}

\renewcommand{\vec}[1]{\mathbf #1}

\newcommand{\gam}{\gamma}

\newcommand{\kap}{\kappa}
\newcommand{\lam}{\lambda}
\newcommand{\vhi}{\varphi}
\newcommand{\sig}{\sigma}

\newcommand{\id}{\mathbf 1}

\newcommand{\x}{\vec r}
\newcommand{\X}{\vec R}
\newcommand{\z}{\{\x_k\}}
\newcommand{\Da}{D_\text{a}}
\newcommand{\tx}{\tau_\text{r}}
\newcommand{\nois}{\boldsymbol\xi}
\newcommand{\mmu}{\boldsymbol\mu}

\newcommand{\kT}{k_\text{B}T}

\begin{document}

\title{Geometric view of stochastic thermodynamics for non-equilibrium steady states}

\author{Thomas Speck}
\affiliation{Institut f\"ur Physik, Johannes Gutenberg-Universit\"at Mainz,
  Staudingerweg 7-9, 55128 Mainz, Germany}

\begin{abstract}
  We explore the idea that non-equilibrium steady states breaking detailed balance are obtained by deforming trajectories (lines in space-time) that have been sampled in a reference system with stochastic dynamics obeying detailed balance, and we ask for the work required to perform this task. These geometric deformations are not arbitrary but arise through interactions with the environment, either the manipulation of conserved quantities by an external agent, or by their exchange with a work reservoir. This view allows to consistently model the breaking of detailed balance and the accompanying entropy production without non-conservative forces, and to systematically extend the notion of thermodynamic ensembles to non-equilibrium steady states. We illustrate the usefulness of this approach by applying it to suspensions of active colloidal particles and deriving their thermodynamically consistent equations of motion.
\end{abstract}


\maketitle


\section{Introduction}

Any quantitative numerical prediction (such as rates, phases, binding energies, \emph{etc.}) across chemistry, biology, and physics starts with a model: the relevant degrees of freedom endowed with an Hamiltonian. Solving the (typically classical) equations of motion samples configurations and trajectories from this model, which are compatible with the microcanonical ensemble preserving the value of the Hamiltonian. Different environmental constraints (such as constant pressure versus constant volume) correspond to different statistical ensembles~\cite{chandler}. Correct sampling from these ensembles is typically achieved through the \emph{extended ensemble approach} of molecular dynamics~\cite{frenkel} pioneered by Andersen~\cite{ande80}, in which the state space of the system is extended by (effective) degrees of freedom modelling the interactions with the environment. This has proven to be an immensely powerful technique that allows to numerically predict material properties at a variety of external conditions. However, it is restricted to thermal equilibrium and so far no systematic extension to non-equilibrium states has been provided.

At thermal equilibrium, the microscopic dynamics governing the motion of particles obeys \emph{detailed balance}, a condition that guarantees the absence of directed transport (no preferred direction, no currents) and a vanishing entropy production. The fundamental symmetry is time-reversal: in equilibrium we cannot distinguish whether a movie is played forward or backward. This is different in driven systems, with the dissipation rate determining time asymmetry~\cite{feng08} and bounding uncertainties~\cite{bara15,ging16}. Consequently, understanding how detailed balance is broken in driven systems is pivotal for consistent and accurate modeling.

The purpose of this manuscript is to show how extended ensembles can be constructed for driven non-equilibrium steady states. Exploiting concepts and insights from stochastic thermodynamics~\cite{seif12}, our approach yields equations of motion that are thermodynamically consistent by construction. While, \emph{e.g.}, non-conservative fields explicitly breaking detailed balance have been considered extensively, here we are primarily interested in more complex systems with (conformational) changes that are driven mechanically and by converting chemical energy. Our strategy can be applied to any existing model upon identifying the \emph{geometric deformation} caused by exchanges with the environment. In Sec.~\ref{sec:motiv}, we will illustrate the basic idea for a sheared colloidal suspension after recapitulating the statistical foundation of barostats. In Sec.~\ref{sec:general} the general formalism is developed.

To demonstrate the power of this approach, in Sec.~\ref{sec:active} we will apply it to a model for active particles (cellular~\cite{pros15} and colloidal~\cite{bech16}), which are characterized by their directed motion. While particles move autonomously (no external guiding field), there is a preferred direction that evolves in time. Even in the absence of particle currents the dynamics breaks detailed balance, implying a non-vanishing heat dissipation. In experiments on active colloidal particles, the energy to sustain the directed motion is supplied locally, typically through light~\cite{jian10,butt12} or chemically through the decomposition of hydrogen peroxide~\cite{paxt04,hows07,pala10}. Such active particles have become the focus of intensive research due to, among many other reasons, novel collective behavior like motility induced phase separation in the absence of attractive forces~\cite{thom11,yaou12,butt13,bial14,cate15} and possible applications in the self-assembly of colloidal materials~\cite{kumm15,meer16}. Active suspensions have already been exploited to power microscale devices~\cite{soko09,leon10,kais14,krish16}, for templated self-assembly~\cite{sten16}, and to set up spontaneous flows on macroscopic lengths~\cite{wu17}. For simple models of active particles the entropy production has been studied recently but with conflicting definitions and results~\cite{gang13,fala16a,fodo16,spec16,mand17,marc17,piet17}. We will show that active colloidal particles share similarities with molecular motors and sheared suspensions. Consequently, the same framework of stochastic thermodynamics can be applied, providing an unambiguous and physically transparent identification of work and heat.

Stochastic thermodynamics applies to systems in contact with an ideal heat reservoir that provides equilibrium-like fluctuations even when the system is strongly driven away from equilibrium. It has been tested experimentally, \emph{e.g.}, for the mechanical unfolding of RNA~\cite{coll05} and a colloidal particle driven by laser tweezers~\cite{blic06}. Initially, stochastic thermodynamics has evolved in response to Jarzynski's and Crook's seminal work relations~\cite{jarz97,croo99}, in which an external agent manipulates some parameter (such as the position of the laser tweezers) to drive the system from initial to final state. The dynamics is non-autonomous and obeys detailed balance, with dissipation due to the work spent by the external agent. However, many systems are driven autonomously (without external interference) through coupling their boundaries to environments with different temperatures, chemical potentials, \emph{etc.}; forcing currents through the system. A cornerstone of the extension of stochastic thermodynamics to such systems~\cite{schm07,broeck15} is the \emph{local detailed balance} condition, which relates the dissipation measured from the autonomous dynamics to these currents. Since then further generalizations of the mechanism how systems are driven have been discussed, in particular driving through feedback~\cite{toya10,rosin15} and information reservoirs~\cite{horo14,barato14}.


\section{Background and motivation}
\label{sec:motiv}

\subsection{Stochastic thermodynamics}

The conventional basis of stochastic energetics is the distinction between degrees of freedom that are controlled (in the following denoted by the vector $\vec X$) and stochastic degrees of freedom that evolve under the influence of thermal noise~\cite{seki98,sekimoto}. Throughout, we will confine our discussion to a suspension of $N$ spherical colloidal particles with positions $\z$ moving in a solvent at constant temperature $T$. We assume a scale separation so that on the time scale the positions change the other degrees of freedom (momenta and solvent degrees of freedom) have equilibrated. Integration over these equilibrated degrees of freedom yields the free energy $H(\z;T,\vec X)=\mathcal F_\text{id}(T,\vec X)+U(\z;\vec X)$ with potential energy $U(\z;\vec X)$ of the colloidal particles. Moreover, we assume the weak coupling regime, \emph{i.e.}, the ideal contribution $\mathcal F_\text{id}$ (including the solvent) does not depend on the positions of the colloidal particles.

A change of the free energy
\begin{equation}
  \label{eq:first}
  \td{H}{t} = \pd{H}{\x_k}\cdot\dot\x_k + \pd{H}{X_i}\dot X_i 
  = -\dot q + \dot w_\text{ag}
\end{equation}
is either work $w_\text{ag}$ (due to the agent manipulating the quantities $\vec X$) or heat $q$ due to a change of the positions. Throughout, the dot denotes a rate whereas total derivatives are $\td{}{t}$. Eq.~(\ref{eq:first}) is the first law describing the conservation of energy along every stochastic trajectory of the system characterized by the (in principle measurable) positions of all particles.

So far, the work spent on the system is due to an external (idealized) agent who can precisely control $\vec X$ (\emph{e.g.}, the position of a laser trap). Now suppose the work is not provided by an agent but removed from a work reservoir, in the simplest case a weight that can be lowered thus liberating potential energy. What is the relation between these two \emph{ensembles}, the one generated by an agent and the ensemble generated by draining the work reservoir? A steady state is reached both for constant \emph{fluxes} $\dot{\vec X}$ and a work reservoir characterized by constant \emph{affinities} $\vec f$. To make contact with conventional thermodynamics, we require that $X_i\leftrightarrow f_i$ are conjugate quantities and that, moreover, the (free) energy of the work reservoir can be expressed as
\begin{equation}
  \label{eq:G:res}
  G_\text{res} = G_0 - f_iX_i
\end{equation}
with constant $G_0$. Here and in the following we employ the sum convention and sum over repeated indices. We will show that both ensembles are related through a geometric connection based on a mapping $\x_k\to\X_k(\x_k,\vec X)$ of particle positions such that the potential energy
\begin{equation}
  U(\z;\vec X) = U(\{\X_k\})
\end{equation}
only depends on the transformed positions.

\subsection{Extended ensemble approach}

\begin{figure}[b!]
  \centering
  \includegraphics{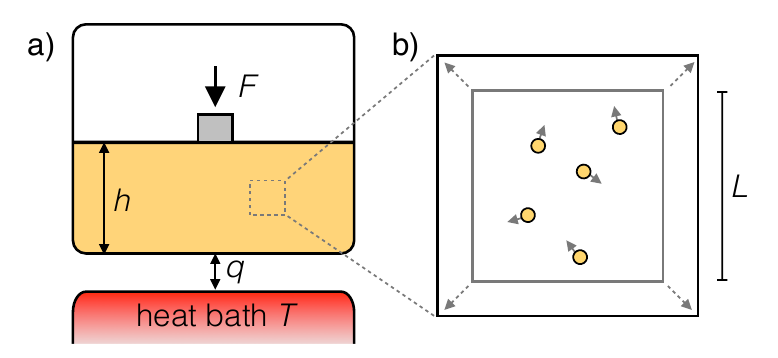}
  \caption{(a)~A suspension in contact with a heat reservoir at temperature $T$ and confined by a movable piston. Equilibrium is reached when the external force is balanced by the (average) pressure of the suspension. (b)~To calculate the pressure in simulations, we consider a small subsystem with volume $V=L^d$ containing $N$ interacting particles with potential energy $U$. A small change $\delta V$ of the volume can be modeled as a uniform rescaling of particle positions, which changes the potential energy. The reversible work for this (virtual) change then is $-p\delta V$ yielding the pressure $p$.}
  \label{fig:pressure}
\end{figure}

The idea of deforming a reference system to sample an \emph{equilibrium} statistical ensemble is also exploited in the extended ensemble approach, notably Andersen's barostat~\cite{ande80} and the generalization to shape changes by Parrinello and Raman~\cite{parr81}. To recall, let us consider a vessel containing a suspension confined to the lower part by a movable piston [Fig.~\ref{fig:pressure}(a)]. The piston is hold down by a weight (we ignore the influence of gravity on the suspension and assume hard walls).
The total (system plus work reservoir) free energy of the vessel is $H_\text{tot}=H+G_\text{res}$ and thus
\begin{equation}
  H_\text{tot}(\z,V;T,p) = \mathcal F_\text{id}(T,V) + U(\z;V) + Fh,
\end{equation}
where $h$ is the height of the piston and $F$ is the force due to the weight. With area $A$ of the piston and mechanical pressure $p=F/A$ we have $Fh=pV$ with volume $V=Ah$. Now $p$ is fixed but $V$ can change due to thermal fluctuations; there is an exchange of volume between suspension and the remainder of the vessel acting as a volume reservoir with potential energy $G_\text{res}=pV$. Assuming that the vessel has settled to an equilibrium state, these fluctuations are governed by the \emph{joint} Boltzmann distribution
\begin{equation}
  \label{eq:boltz:p}
  \Psi_\text{eq}(\z,V) \propto e^{-\beta H_\text{tot}}
\end{equation}
of reference positions and volume. As usual, we denote $\beta\equiv(\kT)^{-1}$ the inverse thermal energy with Boltzmann's constant $k_\text{B}$. This is the well-known result from statistical mechanics for a system at constant pressure.

What makes this approach applicable in computer simulations is that Eq.~(\ref{eq:boltz:p}) still holds in a small subvolume (but larger than the correlations length) as depicted in Fig.~\ref{fig:pressure}(b). Let this subbox be cubic and its volume be $V=L^d$ with dimension $d=3$. Choosing stochastic dynamics, the evolution of the volume obeying Eq.~(\ref{eq:boltz:p}) is
\begin{equation}
  \label{eq:V}
  \dot V = -\Gamma\pd{(\beta H_\text{tot})}{V} + \zeta_V
  = -\Gamma\beta[p-\hat p(\z;V)] + \zeta_V
\end{equation}
with arbitrary mobility $\Gamma$ (but independent of $\z$) and Gaussian noise $\zeta_V$ having correlations $\mean{\zeta_V(t)\zeta_V(s)}=2\Gamma\delta(t-s)$. The derivative becomes
\begin{equation}
  \hat p = -\pd{H_\text{tot}}{V} = p_\text{id} + \hat p_\text{i}
\end{equation}
with contribution $p_\text{id}(T,V)=-\pd{\mathcal F_\text{id}}{V}$ of the equilibrated degrees of freedom, which is a sum of the ideal gas pressure of the colloidal particles and the pressure due to the solvent. The contribution to the pressure due to particle interactions can be expressed as an equilibrium average over the instantaneous pressure
\begin{equation}
  \label{eq:p:hat}
  \hat p_\text{i}(\z;V) = -\pd{U}{V}
\end{equation}
depending on the microstate.

To see how the mapping enters, consider a cubic unit box with reference positions $\z$ and employing periodic boundary conditions. The actual, ``deformed'' positions are then $\X_k=L\x_k$, which thus are functions of the volume. The partition function reads
\begin{equation}
  Z(T,V) = \sum_{\z} e^{-\beta H(\z;V)} = e^{-\beta\mathcal F(T,V)}
\end{equation}
with free energy $\mathcal F(T,V)$. An infinitesimal change of the volume $\delta V$ (at constant temperature) requires the work $\delta w=\dd\mathcal F=-\mean{\hat p}\delta V$, which in equilibrium is reversible and given by the change of free energy. Hence, the partial derivative of the free energy with respect to volume is the (negative) pressure
\begin{equation}
  \label{eq:pressure}
  -\mean{\hat p} = \pd{\mathcal F}{V} = -p_\text{id} 
  + \sum_{\z} \pd{U}{V} \psi_\text{eq}
  = -p_\text{id} - \mean{\hat p_\text{i}}
\end{equation}
with Boltzmann distribution
\begin{equation}
  \label{eq:boltz:ref}
  \psi_\text{eq}(\z;T,V) \propto e^{-\beta H(\z;T,V)}
\end{equation}
of the reference system. Throughout, the brackets $\mean{\cdot}$ denote an average. We, therefore, find that the explicit deformation of the simulation box yields the same expression for the pressure that appears in the evolution of the volume Eq.~(\ref{eq:V}). With $\mean{\dot V}=0$ we immediately find $p=\mean{\hat p}$, \emph{i.e.}, the average pressure in the suspension is the same as the mechanical pressure maintained by the volume reservoir. The purpose of this work is to demonstrate how this extended ensemble approach can be applied to non-equilibrium states.

\subsection{Simple shear}
\label{sec:strain}

\begin{figure}[b!]
  \centering
  \includegraphics{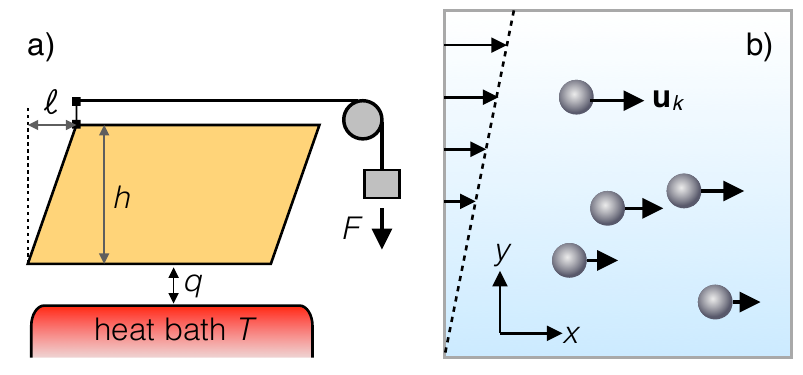}
  \caption{(a)~Sheared system coupled to a constant stress reservoir. The upper wall is displaced by $\ell$ with shear strain $\gam=\ell/h$. The reservoir is modeled as a weight [cf. Fig.~\ref{fig:pressure}(a)]. (b)~Colloidal suspension in simple shear flow with local solvent velocity $\vec u_k$.}
  \label{fig:shear}
\end{figure}

\subsubsection{Constant stress reservoir}

For an illustration of an extended ensemble in non-equilibrium, we consider an elastic solid attached to walls with the geometry shown in Fig.~\ref{fig:shear}(a). Initially a weight is lifted a distance $L$ that exerts a shear force $F\vec e_x$ on the movable upper wall with area $A$, which is displaced by $\ell$. The volume of the solid is $V=Ah$ independent of $\ell$. The shear stress is $\sig=F/A$ and the potential energy of the work reservoir becomes
\begin{equation}
  \label{eq:res:strain}
  G_\text{res} = F(L-\ell) = FL - \sig V\gam
\end{equation}
with $G_0=FL$ the initial, reversible work to create the reservoir (lift the weight) and $\gam\equiv\ell/h$ the shear strain.

Now suppose we replace the solid by a liquid (or suspension) that cannot sustain any shear stress and starts to flow. The upper wall will move with non-zero average speed and (ignoring the possibility of shear banding~\cite{olmsted08}) a linear flow profile will be observed. The weight is steadily lowered and the potential energy of the reservoir is reduced. This energy is eventually dissipated into the heat bath due to the viscosity of the solvent. We can easily quantify this dissipated heat with rate
\begin{equation}
  \label{eq:first:res}
  \dot q_\text{tot} = -\td{H_\text{tot}}{t} = \dot w_\text{res} - \td{H}{t}
\end{equation}
since any change of the total energy $H_\text{tot}(\z,\gam)$ is necessarily due to an exchange of energy with the heat bath. In the second step, we have inserted the total energy with
\begin{equation}
  \dot w_\text{res} = -\td{G_\text{res}}{t},
\end{equation}
\emph{i.e.}, the energy transfered from work reservoir to the suspension is identified as work (exerted by the reservoir on the system). Specifically, from Eq.~(\ref{eq:res:strain}) we obtain $\dot w_\text{res}=\sig V\dot\gam$, which can be integrated to the work $w_\text{res}=\sig V\Delta\gam_t$ with $\Delta\gam_t$ the change of strain over a given observation time $t$. The work thus has the expected bilinear form of an intensive affinity (the stress $\sig$) times the change of the conjugate extensive quantity ($V\gam$).

\subsubsection{Deformation}
\label{sec:deform}

Considering a microscopic sample volume as before, the shape of the target system can be obtained through deforming a reference system according to $\X_k=\vec h\cdot\x_k$ with matrix
\begin{equation}
  \label{eq:h}
  \vec h = \left(
    \begin{array}{ccc}
      1 & \gam & 0 \\ 0 & 1 & 0 \\ 0 & 0 & 1
    \end{array}\right), \quad
  \vec h^{-1} = \left(
    \begin{array}{ccc}
      1 & -\gam & 0 \\ 0 & 1 & 0 \\ 0 & 0 & 1
    \end{array}\right)
\end{equation}
following the same route as for the pressure (in that case $\vec h=V^{1/d}\id$). The work for an infinitesimal change of the strain becomes $\delta w=\hat\sig V\delta\gam$ with the instantaneous off-diagonal shear stress
\begin{equation}
  \label{eq:stress}
  \hat\sig(\z;\gam) = \pd{H}{(V\gam)} = \frac{1}{V}\sum_{k=1}^Ny_k\pd{U}{x_k}
\end{equation}
calculated from the particle configuration. In equilibrium one finds $\mean{\hat\sig}=\sig$.

\subsubsection{Stochastic energetics}

Modeling a colloidal suspension as shown in Fig.~\ref{fig:shear}(b) driven into a non-equilibrium steady state by (freely draining) simple shear flow with constant strain rate $\dot\gam$, one would write down the evolution equations (in target space)
\begin{equation}
  \label{eq:shear}
  \dot\X_k = \vec u_k - \mu_0\pd{U}{\X_k} + \nois_k,
\end{equation}
where $\vec u_k=\dot\gam Y_k\vec e_x$ is the solvent velocity. The Gaussian white noises have zero mean and correlations $\mean{\nois_k(t)\nois^T_l(s)}=2D_0\delta_{kl}\mathbf 1\delta(t-s)$ with strength $D_0=\kT\mu_0$, where $T$ is the temperature of the solvent (acting as the heat bath) and $\mu_0$ is the bare Stokes mobility. The entropy production rate $\dot{\mathcal S}$ (as calculated from time reversal, see appendix~\ref{sec:reversal}) reads
\begin{equation}
  \frac{\dot{\mathcal S}}{\beta} 
  = -\left(\dot\X_k-\vec u_k\right)\cdot\pd{U}{\X_k}
  \overset{!}{=} \dot q
\end{equation}
with total derivative $\dot\X_k\cdot\pd{U}{\X_k}=\td{U}{t}$. Thermodynamic consistency requires to identify the dissipated heat $\dot q$ with the entropy produced in the heat bath. Exploiting the first law Eq.~(\ref{eq:first}), we identify the work rate
\begin{equation}
  \label{eq:w:shear}
  \dot w_\text{ag} = \vec u_k\cdot\pd{U}{\X_k} = \hat\sig(\{\X_k\})V\dot\gam
\end{equation}
spent by an external agent to maintain the non-equilibrium steady state. This expression is in agreement with the work due to an explicit small change of the strain (previous Sec.~\ref{sec:deform}). The same work rate is obtained from general considerations on the invariance of work and heat with respect to the frame of reference~\cite{spec08}. Moreover, this expression has the same form as the reservoir work rate but with the stress replaced by the instantaneous stress Eq.~(\ref{eq:stress}) in the suspension. Note that Eq.~(\ref{eq:w:shear}) only accounts for the work spent against the external flow and not the work required to generate the flow.


\section{General formalism}
\label{sec:general}

\subsection{Reference system}
\label{sec:ref}

We now generalize the results of the previous section. Our starting point is a reference system in thermal equilibrium. For concreteness, we consider $N$ colloidal particles moving in an aqueous solvent with coupled equations of motion
\begin{equation}
  \label{eq:lang}
  \dot\x_k = -\mu_0\pd{U}{\x_k} + \nois_k,
\end{equation}
where mobility $\mu_0$ and noise $\nois_k$ are as in Eq.~(\ref{eq:shear}). It is well established that this dynamics obeys detailed balance and samples the microstates $\z$ according to the Boltzmann distribution Eq.~(\ref{eq:boltz:ref}) with potential energy $U(\z)$. In Eq.~(\ref{eq:lang}), we have neglected hydrodynamic coupling between particles due to the solvent. While such a coupling strongly influences the dynamics, it does not change the dissipation nor the expressions for work discussed in the following (for details see appendix~\ref{sec:hydro}).

The second ingredient is the ``deformation''
\begin{equation}
  \label{eq:map}
  \mathcal T : \x_k \mapsto \X_k(\x_k,\vec X)
\end{equation}
moving particles to new positions $\X_k$ that depend on $m$ additional variables $\vec X=(X_1,\dots,X_m)$. Clearly, such a deformation will require (release) work to move particles against (with) the potential energy. Formally, $\mathcal T$ describes a mapping of positions onto new positions parametrized by $\vec X$, and we require throughout that the inverse mapping $\mathcal T^{-1}$ exists with $\mathcal T\circ\mathcal T^{-1}=1$. Moreover, we restrict our attention to mappings that keep the volume constant, which implies a Jacobian determinant with value 1.

In the following, we will extend this procedure of deforming particle positions depending on (conserved) quantities $\vec X$ to describe non-equilibrium steady states. The microscopic dynamics of the reference system obeys detailed balance so that, holding $\vec X$ fixed, its steady state corresponds to thermal equilibrium at inverse temperature $\beta$.

\subsection{Constant-flux ensemble}
\label{sec:flux}

In analogy with the instantaneous pressure [Eq.~(\ref{eq:p:hat})], we introduce the conjugated, \emph{instantaneous} forces
\begin{equation}
  \label{eq:conj}
  \hat f_i(\z;\vec X) \equiv \pd{H}{X_i} = f^{(i)}_\text{id} +
  \vec d^{(i)}_k\cdot\pd{U}{\x_k}
\end{equation}
so that the work rate takes the bilinear form $\dot w_\text{ag}=\hat f_i\dot X_i$. We have applied the chain rule to rewrite the partial derivative of the potential, which defines the effective displacements $\vec d^{(i)}_k(\z;\vec X)$ of particles describing the deformation. For the examples in Sec.~\ref{sec:motiv}, we find $\vec d^{(V)}_k=\x_k/(dV)$ for the volume and $\vec d^{(\gam)}_k=y_k\vec e_x$ for the strain. This decomposition of the conjugate thermodynamic forces into mechanical forces in the reference system and displacements $\vec d^{(i)}_k$ is our first main result.

Treating the mapping Eq.~(\ref{eq:map}) as a variable transformation, the stochastic dynamics in target space would read (see, \emph{e.g.}, Ref.~\cite{risken})
\begin{equation}
  \label{eq:lang:trafo}
  \dot\X_k = \td{\X_k}{t} = \pd{\X_k}{X_i}\dot X_i + \pd{\X_k}{\x_k}\cdot\dot\x_k.
\end{equation}
However, following this dynamics the entropy production (calculated through time reversal, see appendix~\ref{sec:reversal}) is different from the heat identified through the first law [Eq.~(\ref{eq:first})]. The corresponding steady state is thus different from the steady state reached through enforcing constant fluxes $\dot X_i$ in the reference system. To obtain exactly the same dissipation, we need to employ the dynamics
\begin{equation}
  \label{eq:lang:const}
  \dot\X_k = \pd{\X_k}{X_i}\dot X_i - \mu_0\pd{U}{\X_k} + \nois_k
\end{equation}
with the same noise statistics as in Eq.~(\ref{eq:lang}), see the structure of Eq.~(\ref{eq:shear}). Now the asymmetric term under time reversal reads
\begin{equation}
  \begin{split}
    \frac{\dot{\mathcal S}}{\beta} &=
    -\left(\dot\X_k-\pd{\X_k}{X_i}\dot X_i\right)\cdot\pd{U}{\X_k} \\
    &= -\td{U}{t} + \pd{U}{X_i}\dot X_i
    = -\td{H}{t} + \dot w_\text{ag} = \dot q
  \end{split}
\end{equation}
inserting on the second line the conjugate forces Eq.~(\ref{eq:conj}) and then the first law Eq.~(\ref{eq:first}). We stress that this disagreement of Eq.~(\ref{eq:lang:trafo}) with Eq.~(\ref{eq:lang:const}) is a consequence of the transformation being an ``active'' deformation moving particles (and has already been noted by Andersen~\cite{ande80}).

Eq.~(\ref{eq:lang:const}) describes the autonomous dynamics of a class of systems which are driven by a non-potential ``flow'' $\vec u_k=\pd{\X_k}{X_i}\dot X_i$ breaking detailed balance. Typically, one would start by writing down this equation. Going backwards, what we have thus demonstrated is that there is a decomposition of the positions $\X_k$ into reference positions $\x_k$ and (conserved) quantities $\vec X$ such that the reference positions are governed by a dynamics that obeys detailed balance with respect to the same potential energy $U(\{\X_k\})=U(\z;\vec X)$.

\subsection{Constant-affinity ensemble}
\label{sec:affinity}

\begin{figure}[t]
  \centering
  \includegraphics{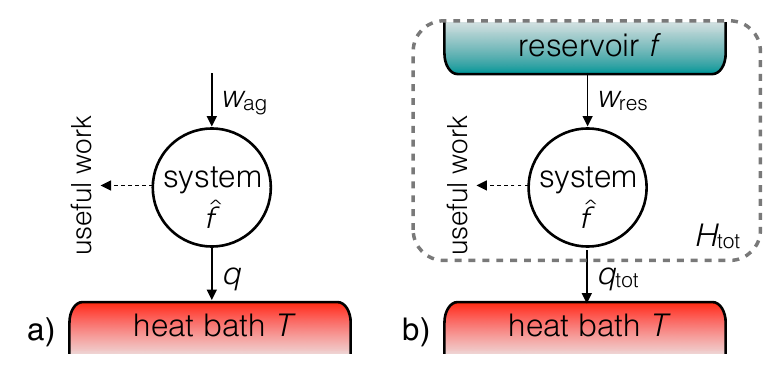}
  \caption{Non-equilibrium ensembles. (a)~Constant-flux ensemble. An external agent manipulates a parameter $X$ with constant rate. This requires the work $w_\text{ag}$ and drives the system into a non-equilibrium steady state. The system is characterized by the conjugate forces $\hat f(\z;X)$ depending on the system's microstate $\z$. Useful work might be extracted, the remaining work is dissipated as heat $q$ into a heat bath at temperature $T$ increasing the entropy of the universe by $\beta q$. (b)~Constant-affinity ensemble. Instead of an agent, the system is now coupled to an ideal work reservoir (characterized by affinity $f$), both of which form a super-system (dashed boundary) with total energy $H_\text{tot}$. A non-equilibrium steady state is reached for $\mean{\hat f}\neq f$ continuously lowering the energy of the reservoir, which is spent as work $w_\text{res}$ on the system and eventually dissipated into the heat bath.}
  \label{fig:ensembles}
\end{figure}

We now replace the agent driving the system by a work reservoir with which the system exchanges the same variables $\vec X$ that were manipulated externally in the constant-flux ensemble, see schematic of Fig.~\ref{fig:ensembles}. The reservoir is assumed to be described by the (free) energy Eq.~(\ref{eq:G:res}), where the conjugated affinities $\vec f$ are a property of the reservoir and to be distinguished from the instantaneous forces defined in Eq.~(\ref{eq:conj}). The stochastic thermodynamics follows as described in Sec.~\ref{sec:strain} with bilinear reservoir work rate $\dot w_\text{res}=f_i\dot X_i$.

For vanishing fluxes $\mean{\dot X_i}=0$, the combined system plus work reservoir reaches thermal equilibrium with joint probability
\begin{equation}
  \label{eq:boltz}
  \Psi_\text{eq}(\z,\vec X) \propto e^{-\beta H_\text{tot}} = e^{-\beta[U(\z;\vec X)-f_iX_i]}
\end{equation}
given by the Boltzmann factor. The situation we are interested in is when the flux between system and reservoir is non-zero and the system ``drains'' the reservoir. It thus performs work $\dot w_\text{res}$ on the system that decreases its energy $G_\text{res}$. While the system is driven with a time-dependent joint probability $\Psi(\z,\vec X;t)$ different from Eq.~(\ref{eq:boltz}), the dynamics of the combined system-reservoir still obeys detailed balance. We assume that the work reservoir is ideal, \emph{i.e.}, even though the $\vec X$ change over time we assume that the conjugated affinities $\vec f$ remain constant. This assumption is of course an idealization and at some point will break down. Still, as long as it (approximately) holds, the system is in a non-equilibrium steady state. While this is a natural formulation of a steady state, its consequences have not yet been explored in the context of stochastic thermodynamics (with the notable exception of Ref.~\cite{horo16}).

In the extended state space $(\z,\vec X)$, the quantities $\vec X$ also become random variables that fluctuate due to the coupling with the heat bath. For quantities $X_i$ that take continuous values, the equations of motion read
\begin{equation}
  \dot X_i = -\Gamma\pd{(\beta H_\text{tot})}{X_i} + \zeta_i
\end{equation}
with generalized ``mobility'' $\Gamma$ (which we assume to be constant). The correlations $\mean{\zeta_i(t)\zeta_j(s)}=2\Gamma\delta_{ij}\delta(t-s)$ of the noise are again dictated by the fluctuation-dissipation theorem. With Eq.~(\ref{eq:conj}) we thus obtain the evolution equation
\begin{equation}
  \label{eq:Xdot}
  \dot X_i = \Gamma\beta[f_i-\hat f_i(\z;\vec X)] + \zeta_i,
\end{equation}
which shows that the fluxes $\mean{\dot X_i}=0$ vanish for $\mean{\hat f_i}=f_i$, \emph{i.e.}, uniform conjugate forces for system and reservoir as expected for equilibrium. Conversely, transport of $X_i$ is caused by a difference of $f_i$ and $\mean{\hat f_i}$ between reservoir and system.

Writing down the Fokker-Planck equation for the stochastic process described by Eqs.~(\ref{eq:lang}) and~(\ref{eq:Xdot}), we obtain
\begin{multline}
  \label{eq:fp}
  \partial_t\Psi = \mu_0\nabla_k\cdot[(\nabla_kU)+\beta^{-1}\nabla_k]\Psi \\
  + \Gamma\pd{}{X_i}\left[-\beta(f_i-\hat f_i)
    +\pd{}{X_i}\right]\Psi
\end{multline}
for the joint distribution $\Psi(\z,\vec X;t)$. It is straightforward to check that the Boltzmann distribution Eq.~(\ref{eq:boltz}) is the stationary solution of Eq.~(\ref{eq:fp}), which is independent of the mobilities. In a non-equilibrium steady state, the joint probability $\Psi(\z,\vec X;t)$ remains explicitly time-dependent but with the marginal distribution $\Int{\vec X}\Psi$ being time-independent.

\subsection{Discrete state space}
\label{sec:discrete}

For completeness, we also discuss the case of a discrete state space $\{\z\}$. In this case, the dynamics is described by jump rates. For the combined system-reservoir, the microscopic jump rates obey the detailed balance condition
\begin{align}
  \label{eq:db}
  \frac{\kap(\z\to\z',\vec X\to\vec X+\delta\vec X)}{\kap(\z'\to\z,\vec X+\delta\vec X\to\vec X)} 
  &= \frac{\Psi_\text{eq}(\z',\vec X+\delta\vec X)}{\Psi_\text{eq}(\z,\vec X)} \\
  &= e^{-\beta[\delta U-f_i\delta X_i]}
\end{align}
with respect to the Boltzmann distribution Eq.~(\ref{eq:boltz}) even if a non-equilibrium steady state is reached due to the exchange of $\vec X$. This is known as \emph{local detailed balance}. Here, $\delta U$ and $\delta\vec X$ are the change of potential energy and the extensive quantities in the transition, respectively. Taking the logarithm of Eq.~(\ref{eq:db})
\begin{equation}
  \kT\ln\frac{\kap(\z\to\z')}{\kap(\z'\to\z)} = -\delta U + f_i\delta X_i 
  = \delta q_\text{tot}
\end{equation}
and appealing to the first law with reservoir work $\delta w_\text{res}=f_i\delta X_i$ leads to the identification of the ratio of forward to backward transition with the heat $\delta q_\text{tot}$ dissipated during this transition, cf. Eq.~(\ref{eq:first:res}).

In the constant-flux ensemble of systems with discrete state space, we split the change of potential energy $\delta U=\delta U|_{\vec X}+\delta U_{\z}$ into a contribution holding the $\vec X$ fixed and a contribution at fixed microstate $\z$ externally changing $\vec X$ [cf. Eq.~(\ref{eq:first})]. The latter is identified with the work
\begin{equation}
  \label{eq:w:discrete}
  \delta w_\text{ag} = \delta U|_{\z} \approx \hat f_i\delta X_i,
\end{equation}
where we have to assume that the increments $\delta X_i$ are small to allow for an expansion of the potential energy with conjugate force Eq.~(\ref{eq:conj}). The second term then is the dissipated heat
\begin{equation}
  \delta q = -\delta U|_{\vec X} 
  = \kT\ln\frac{\kap_\text{ref}(\z\to\z')}{\kap_\text{ref}(\z'\to\z)},
\end{equation}
which can be related to the jump rates of the reference system since we had assumed that these fulfil detailed balance (at fixed $\vec X$).

\subsection{Equilibrium}

The well-known thermodynamic relations between intensive ($f_i$) and extensive ($X_i$) quantities are recovered when considering a small, instantaneous perturbation of the equilibrium reference state through small changes $\delta X_i$ requiring the work $\delta w=\hat f_i\delta X_i$. This situation can be regarded as a \emph{virtual} perturbation since the average work $\mean{\hat f_i}_\text{eq}\delta X_i$ is calculated with respect to the equilibrium Boltzmann distribution so that
\begin{equation}
  \mean{\hat f_i}_\text{eq} = \sum_{\z} \pd{U}{X_i}\psi_\text{eq}(\z) 
  = \pd{\mathcal F}{X_i}
\end{equation}
with the free energy $\mathcal F(T,\vec X)$, which generalizes Eq.~(\ref{eq:pressure}). Hence, perturbing equilibrium, the average conjugate forces can be related to derivates of the thermodynamic potential.

In the constant-affinity ensemble, in case the Boltzmann distribution Eq.~(\ref{eq:boltz}) is a solution of Eq.~(\ref{eq:fp}) one immediately finds
\begin{equation}
  \mean{X_i}_\text{eq} = \sum_{\z,\vec X} X_i\Psi_\text{eq}(\z,\vec X)
  = -\pd{\mathcal G}{f_i}
\end{equation}
with (Gibbs) free energy $\mathcal G(T,\vec f)$, which is the Laplace transform
\begin{equation}
  e^{-\beta\mathcal G} = \sum_{\vec X} e^{-\beta\mathcal F(T,\vec X)+\beta f_iX_i}
\end{equation}
of the (Helmholtz) free energy $\mathcal F(T,\vec X)$.

\subsection{Fluctuation theorems and linear response regime}

Fluctuation theorems express the broken symmetries of path probabilities. In particular time-reversal in driven processes entails the fluctuation theorem for the total entropy production~\cite{seif05a}, which we have ensured is fulfilled. Another useful fluctuation theorem is the transient work relation $\mean{e^{-\beta w_\text{ag}}}=1$ in the constant-flux ensemble with the system initially prepared in thermal equilibrium. Of course, this is nothing more than the Jarzynski relation combined with the fact that the free energy of the reference system is constant due to our restriction to mappings with unit Jacobian determinant. This relation yields the fluctuation-dissipation theorem~\cite{gall96,andr07b}. To this end, consider the work $w_\text{ag}=\IInt{t'}{0}{t}\hat f_i(t')\dot X_i$ with $\hat f_i(t')=\hat f_i(\z(t'),\vec X(t'))$. We pick a finite observation time $t$ so that the steady state has been reached after this time $t$ for small fluxes $\dot X_i$. Expanding the exponential $1-\beta\mean{\hat f_i}\dot X_i+\frac{1}{2}\beta^2\mean{w_\text{ag}^2}+\cdots$ and taking the derivative with respect to $t$, we obtain
\begin{equation}
  \mean{\hat f_i} = \beta
  \IInt{t'}{0}{t}\mean{\hat f_i(t)\hat f_j(t')}_\text{eq}\dot X_j = R_{ij}\dot X_j
\end{equation}
to linear order of the fluxes. This is the fluctuation-dissipation theorem in the linear response regime with symmetric resistances $R_{ij}$~\cite{onsa31}.

The corresponding transient work relation in the constant-affinity ensemble reads
\begin{equation}
  \label{eq:ft:ca}
  \mean{e^{-\beta w_\text{res}}} = 1,
\end{equation}
see appendix~\ref{sec:ft} for the derivation. Following the same line of manipulations as above leads to the average fluxes
\begin{equation}
  \mean{\dot X_i(t)} = \beta
  \IInt{t'}{0}{t} \mean{\dot X_i(t)\dot X_j(t')}_\text{eq} f_j = L_{ij}f_j
\end{equation}
to linear order of the affinities with symmetric conductances $L_{ij}$. Note that the formalism of thermodynamic potentials can be extended to the linear response regime~\cite{onsa31a,palm17}.


\section{Active particles}
\label{sec:active}


\subsection{Preamble: Molecular motor}

\begin{figure}[b!]
  \centering
  \includegraphics{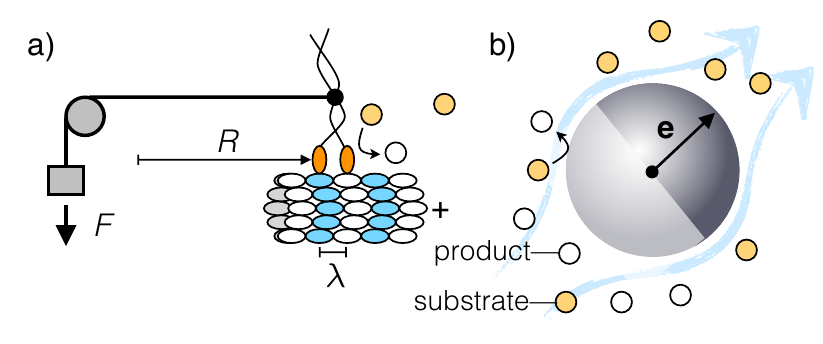}
  \caption{Mechano-chemical coupling. (a)~Sketch of a molecular motor moving along a microtubule and loaded with constant force $F$. Each directed step with step size $\lam$ is due to the conversion of an ATP molecule ($\bullet\to\circ$). (b)~Colloidal Janus particle with two different hemispheres defining its orientation $\vec e$. One hemisphere catalyzes an enzymatic reaction, which depletes the reactants close to this hemisphere. This creates a gradient, which causes a hydrodynamic slip velocity $\vec u$ propelling the particle. We assume that every reaction causes a ``jump'' of the particle along $\vec e$ with step length $\lam$.}
  \label{fig:motor}
\end{figure}

The sheared colloidal suspension in Sec.~\ref{sec:strain} is driven through a non-potential flow that is generated at the boundaries. Another class of driven systems treatable by our approach are systems that are driven through a mechano-chemical coupling, in which mechanical steps are performed due to the free energy released in a chemical reaction. For a simple illustration, consider a single molecular motor (\emph{e.g.} kinesin) moving along a microtubule [Fig.~\ref{fig:motor}(a)]. In equilibrium, the motion is diffusive with zero mean displacement. The motor can be described as an enzyme to which reactant (substrate) molecules $\bullet$ (typically ATP--adenosine triphosphate) bind, hydrolysis of which induces a conformal change that leads to a directed step (with the direction determined by the polarity of the microtubule), after which the product $\circ$ (ADP and P$_\text{i}$) is released. We denote this reaction $\bullet\rightleftharpoons\circ$ with chemical potential difference $\Delta\mu\equiv\mu_\bullet-\mu_\circ$. In addition, the motor performs useful work through lifting a weight.

Following Sec.~\ref{sec:discrete} (see also Ref.~\cite{seif17a}), a consistent set of rates is obtained through combining the motor with reactant and product molecules into a super-system (the ``vessel'') coupled to a heat bath. At the coarsest level of description, we assume tight coupling so that every time a reactant molecule is bound and hydrolyzed the motor performs a step~\cite{juli97,kolo13}. In this limit, $n=n_\circ-n_\circ^0$ is the number of reactions (and thus the number of product molecules produced) during the observation time with $n_\circ^0$ the number of product molecules present at the initial time. Instead of potential energy, the work reservoir now holds the Gibbs free energy
\begin{equation}
  G_\text{res}(n) = \mu_\bullet n_\bullet + \mu_\circ n_\circ = \mu_\bullet n_\text{tot} 
  - n_\circ^0\Delta\mu - n\Delta\mu,
\end{equation}
which takes on the required bilinear form. The total number of molecules $n_\bullet+n_\circ=n_\text{tot}$ remains constant. Clearly, $X\to n$ is the extensive quantity in our formalism, and $f\to\Delta\mu$ corresponds to the driving affinity.

Keeping track of the reactions, we can decompose the actual position $R(r,n)=r+\lam n$ along the microtubule into a reference position that undergoes thermal diffusive steps and $n$ directed steps with constant step length $\lam$ determined by the geometry of the microtubule. The potential energy reads $U(R)=FR$ and the conjugated force Eq.~(\ref{eq:conj}) becomes $\hat f=\partial_nU=\lam F$ with effective displacement $d^{(n)}=\lam$. For tight coupling, $\mean{U}=w_\text{res}=n\Delta\mu$ and the energy of the reservoir is used to increase the potential energy of the system. Equilibrium is reached exactly at the point $F\lam=\Delta\mu$ (the stall force), and for $F>\Delta\mu/\lam$, on average, the motor steps backwards lowering the weight, synthesizing a reactant molecule in every backward step. Note that tight coupling simplifies the description, but more realistic models with several internal states also fall into our framework. These models include idle cycles~\cite{liep07}, during which the reservoir work is dissipated without producing useful work. Formally, we can switch to the constant-flux ensemble with work rate $\dot w_\text{ag}=F\lam\dot n=Fv$ (although in the case of a single molecular motor this seems rather academic as it is not clear how an agent would enforce a constant $\dot n$).

\subsection{Self-propelled colloidal particles}

Instead of a molecular motor, let us now consider a solvated colloidal particle with an inhomogeneous surface, \emph{e.g.}, a spherical Janus particles with one hemisphere coated with a catalyst that promotes a chemical reaction [Fig.~\ref{fig:motor}(b)]. The simplest model is a generic chemical reaction $\bullet\rightleftharpoons\circ$ of a molecular solute~\cite{gole05,gole07}, quite in analogy with the molecular motor. The majority of experimental studies on self-propelled Janus particles exploits the decomposition of hydrogen peroxide (the molecular solute) into hydrogen and oxygen~\cite{bial14,bech16}. An alternative mechanism is the reversible demixing of the molecular solute (specifically, lutidine in water)~\cite{butt12}. In both cases, self-propulsion is powered by the difference in chemical potential (either between reactants and products, or the two phases). Quite generally, an imbalance of local fluxes and mobilities across the particle surface leads to a hydrodynamic slip velocity $\vec u=\lam\dot n\vec e$ due to the reciprocal theorem for Stokes flow~\cite{gole07}. Here, $\vec e$ is the unit orientation of the particle (for a Janus particle it points along the poles of the two hemispheres), $\dot n$ is the total flux of molecular solutes, and $\lam$ is a length that depends on the particle geometry and other specific factors. It is fairly challenging to actually calculate $\lam$, the most common approach being based on the thin boundary layer approximation~\cite{ande89}. For our purposes, however, it is sufficient to retain $\lam$ as a parameter.

Recently, Pietzonka and Seifert have derived a continuous description for active particles inspired by molecular motors, where they start from a discrete lattice with step size $\lam$~\cite{piet17}. In addition to diffusive steps, particles undergoes directed jumps along a random lattice direction $\vec e_k$, consuming a solute molecule in every directed step. In the limit of small $\lam$ (compared to the diameter of the colloidal active particle), the work for each such step can be expanded [cf. Eq.~(\ref{eq:w:discrete})]
\begin{equation}
  \label{eq:conj:ps}
  \delta U|_{\z} = U(\x_k+\lam\vec e_k)-U(\x_k) 
  \approx \lam\vec e_k\cdot\pd{U}{\x_k} = \hat f_k
\end{equation}
(here no summation over the particle index $k$) with effective displacements $\vec d_k=\lam\vec e_k$ along the orientation of each particle. The resulting conjugate force lents itself to the interpretation as the instantaneous chemical potential (difference) of the molecular solutes surrounding colloidal particle $k$.

Without loss of generality, we assume that initially the work reservoir only holds reactant molecules with the total number $n_\text{tot}=n_\bullet+n_\circ$ of molecules remaining constant. The number of product molecules is $n_\circ=\sum_kn_k$ with $n_k$ the number of reactions occurring on the surface of the $k$-th particle. The Gibbs free energy of the reservoir thus is $G_\text{res}(n_\circ)=\mu_\bullet n_\text{tot}-n_\circ\Delta\mu$ with $\vec X=(n_1,\dots,n_k)$. The detailed balance condition Eq.~(\ref{eq:db}) becomes
\begin{equation}
  \label{eq:db:ap}
  \frac{\kap^+_k}{\kap^-_k} = e^{-\beta(\hat f_k-\Delta\mu)}
\end{equation}
with $\kap^+_k$ ($\kap^-_k$) the rate to produce (consume) a product molecule. These rates do not depend on $n_k$. The average solute flux is
\begin{equation}
  \label{eq:ndot}
  \mean{\dot n_k} = \mean{\kap^+_k - \kap^-_k} 
  = \mean{\kap^+_k\left[1-e^{\beta(\hat f_k-\Delta\mu)}\right]}.
\end{equation}
Expanding the exponential for small fluctuations away from $\Delta\mu$, to linear order this result agrees with the Langevin prescription Eq.~(\ref{eq:Xdot}).

By now it should become clear that this model fits into the same framework we have developed for the sheared suspension and molecular motor. The model describes the evolution of particle positions and solute numbers in the constant-affinity ensemble defined by $\Delta\mu$, the difference of chemical potential between reactant and product molecules. As long as $\mean{\dot n_k}>0$, particles undergo directed motion (with the opportunity for $\mean{\dot n_k}<0$ to reverse the solute flux and synthesize reactant molecules~\cite{gasp17}). As for the molecular motor, the actual particle positions
\begin{equation}
  \label{eq:map:ap}
  \X_k(t) = \x_k(t) + \lam\IInt{s}{0}{t} \dot n_k(s)\vec e_k(s)
\end{equation}
can be decomposed into reference positions and the active translations. The latter are history-dependent, summing all discrete displacements with step size $\lam$ along the evolving orientations $\vec e_k$. Still, the effective displacement due to converting one more solute (the partial derivative with respect to $n_k$) is simply $\vec d_k=\lam\vec e_k$ and thus depends only on the current state of the system. We see that the resulting conjugate force becomes [cf. Eq.~(\ref{eq:conj})]
\begin{equation}
  \label{eq:conj:ap}
  \hat f_k(\z;\vec X) = \pd{U}{n_k} = \lam\vec e_k\cdot\pd{U}{\x_k}
\end{equation}
in agreement with Eq.~(\ref{eq:conj:ps}). This application of the geometric approach developed here to self-propelled particles is our second main result.

\subsection{Dynamics in target space}

The evolution of the joint probability $\Psi(\{\X_k\},\{\vec e_k\};t)$ of (actual) particle positions and orientations
\begin{equation}
  \label{eq:fp:ap}
  \partial_t\Psi = \mathcal L_\text{p}\Psi + \mathcal L_\text{a}\Psi
\end{equation}
can be split into a passive part with differential operator $\mathcal L_\text{p}$ and active translations due to the chemical reactions. Assuming these to occur independently, the latter is given by
\begin{multline}
  \label{eq:La}
  \mathcal L_\text{a}^\text{(ca)}\Psi 
  = \sum_{k=1}^N\left\{\kap^+_k\Psi(\X_k-\lam\vec e_k) 
    + \kap^-_k\Psi(\X_k+\lam\vec e_k) \right. \\
  - \left. [\kap^+_k(\X_k+\lam\vec e_k)+\kap^-_k(\X_k-\lam\vec e_k)]\Psi\right\},
\end{multline}
where, for clarity, as arguments we only indicate the particle positions that are shifted.

In order to simplify this expression, we now assume that the potential energy introduces a length scale $\ell$ (typically the size of the particles). Expanding in Eq.~(\ref{eq:La}) the arguments of the joint distribution and the rates to linear order in $\lam/\ell$, we obtain
\begin{equation}
  \label{eq:La:approx}
  \mathcal L_\text{a}^\text{(ca)}\Psi \approx
  -\sum_{k=1}^N\pd{}{\X_k}\cdot[\lam(\kap^+_k-\kap^-_k)\vec e_k\Psi].
\end{equation}
The resulting Langevin equations
\begin{equation}
  \label{eq:abp:v2}
  \dot\X_k = \hat v_k\vec e_k - \mu_0\pd{U}{\X_k} + \nois_k
\end{equation}
thus have acquired a non-linear drift term with speed $\hat v_k\equiv\lam(\kap^+_k-\kap^-_k)$ that breaks detailed balance. Note that in this limit $\lam/\ell\ll1$ there is no active noise from the fluctuations of the chemical events $n_k$, which is negligible compared to the thermal noise of the particle positions.

\subsection{Constant-flux ensemble: Active Brownian particles}

For colloidal Janus particles propelled by the conversion of molecular solutes, the fluxes $\dot n_k$ can be expected to be large and their fluctuations to be small. The corresponding constant-flux ensemble is then realized by enforcing exactly the same constant flux $\dot n_k=\dot n$ of molecular solutes on every Janus particle. The equations of motion are given by Eq.~(\ref{eq:lang:const}) with the non-potential term given by
\begin{equation}
  \label{eq:u:abp}
  \vec u_k = \pd{\X_k}{n_k}\dot n_k = v_0\vec e_k
\end{equation}
with constant speed $v_0\equiv\lam\dot n$ [in contrast to $\hat v_k(\{\X_k\})$ appearing in Eq.~(\ref{eq:abp:v2})]. The resulting model is usually referred to as ``active Brownian particles'' (ABPs). The work spent in order to maintain a constant rate $\dot n$ reads
\begin{equation}
  \label{eq:w:abp}
  \dot w_\text{ag} = \hat f_k\dot n_k = \dot n\lam\vec e_k\cdot\pd{U}{\x_k} 
  = \vec u_k\cdot\pd{U}{\x_k}
\end{equation}
inserting the conjugate forces Eq.~(\ref{eq:conj:ap}). Note that this expression has the same form as the work rate Eq.~(\ref{eq:w:shear}) for the sheared colloidal suspension (remember that from Eq.~(\ref{eq:map:ap}) we have $\pd{}{\X_k}=\pd{}{\x_k}$). 

ABPs have been studied extensively because this model exhibits a non-equilibrium phase transition that resembles liquid-gas phase separation~\cite{tail08}. This transition is reproduced in mean-field theories~\cite{spec15} controlled by the effective speed $v(\rho)$ as a function of the local density $\rho(\x,t)$. From the Langevin equations we obtain $v^\text{(cf)}(\rho)=\mean{\vec e_k\cdot\dot\X_k}=v_0-\mu_0\rho\zeta$, where $\zeta(v_0)$ is the force imbalance coefficient~\cite{bial13} (and using that noise and orientations are uncorrelated, $\mean{\vec e_k\cdot\nois_k}=0$). The effective speed $v^\text{(cf)}$ is thus reduced due to the blocking by other particles~\cite{bial13}. The average work rate per particle
\begin{equation}
  \frac{\mean{\dot w_\text{ag}}}{N} = -\frac{v_0}{\mu_0}(v^\text{(cf)}-v_0) 
  = v_0\rho\zeta \geqslant 0
\end{equation}
represents the frictional loss due to the solvent pushing against slow particles that are blocked (in agreement with the idea of a mechanical ``swim force''~\cite{taka14,yan15}). This work has to be supplied by the external agent to maintain the solvent flow at speed $v_0$.

\begin{figure*}
  \centering
  \includegraphics{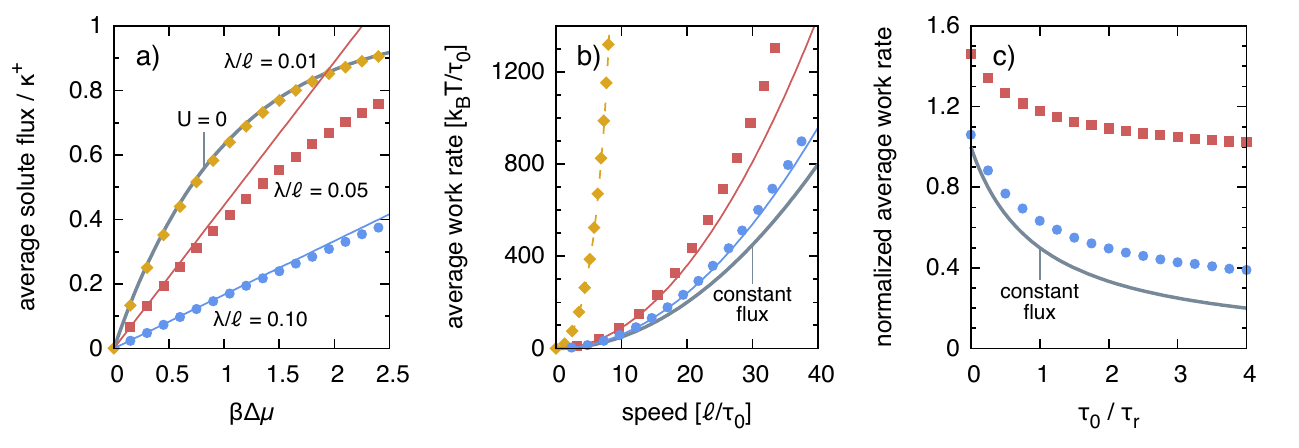}
  \caption{Numerical results for the combined dynamics of particle and reactions $n$ in a harmonic trap. (a)~Average solute flux $\mean{\dot n}$ as a function of $\Delta\mu$ for three lengths $\lam/\ell$ for $\tau_0/\tx=1$ and $\kap^+\tau_0=10^3$. Symbols are numerical results. The solid lines show the linear behavior, see Eq.~(\ref{eq:ndot:lin}). The thick gray line shows the limiting behavior in the absence of a potential ($U=0$). (b)~The average work rate $\mean{\dot w_\text{res}}=\mean{\dot n}\Delta\mu$ as a function of speed $\mean{\hat v}=\lam\mean{\dot n}$. The solid lines show the quadratic behavior Eq.~(\ref{eq:wres:lin}) for small speeds. The dashed line is the limiting behavior for $U=0$ with maximal speed $v_\infty=\lam\kap^+$. The thick gray line shows the work rate Eq.~(\ref{eq:w:cf}) in the constant-flux ensemble. (c)~Average work rate as function of inverse orientation time $\tau_0/\tx$ and normalized by $\mean{\hat v}^2\tau_0/(\beta\ell^2)$. The thick gray line shows the limiting behavior $(1+\tau_0/\tx)^{-1}$.}
  \label{fig:num}
\end{figure*}

For comparison, in the constant-affinity ensemble in the linear regime we find from Eq.~(\ref{eq:abp:v2}) the effective speed
\begin{equation}
  v^\text{(ca)}(\rho) = \lam\kap^+\beta\Delta\mu - (\beta\lam^2\kap^+
  + \mu_0)\rho\zeta.
\end{equation}
The first term is the propulsion speed of free particles. In contrast to ABPs, self-propelled particles in the constant-affinity ensemble are more strongly slowed because the flux of solute molecules [Eq.~(\ref{eq:ndot})] is reduced to compensate for the increased potential energy in denser regions. This agrees qualitatively with experimental results on active colloidal suspensions, which show phase separation already at lower speeds than predicted by ABPs~\cite{butt13}.

\subsection{Illustration: Harmonic trap}

To illustrate the two non-equilibrium ensembles, we now turn to a single active particle moving in two dimensions in the external harmonic potential $U=\tfrac{1}{2}k\X^2$. The stiffness $k$ sets a natural length $\ell\equiv(\beta k)^{-1/2}$ and time scale $\tau_0\equiv(\mu_0k)^{-1}$. The unit orientation $\vec e=(\cos\vhi,\sin\vhi)^T$ is expressed by the angle $\vhi$ it encloses with the $x$-axis. We assume that this orientation undergoes free rotational diffusion with correlation time $\tx$, which leads to the passive evolution operator
\begin{equation}
  \mathcal L_\text{p}\Psi = \frac{1}{\tau_0}\pd{}{\X}\cdot(\X\Psi) 
  + D_0\pd{^2\Psi}{\X^2} + \frac{1}{\tx}\pd{^2\Psi}{\vhi^2}
\end{equation}
in Eq.~(\ref{eq:fp:ap}).

The conjugate force reads $\hat f=k\lam(\vec e\cdot\X)$. The time evolution of the average $\mean{\hat f}$ involving $\Psi$ can be written
\begin{equation}
  \partial_t\mean{\hat f} = 
  -\left(\frac{1}{\tau_0}+\frac{1}{\tx}\right)\mean{\hat f} 
  + k\lam^2\mean{\dot n}
\end{equation}
after inserting Eq.~(\ref{eq:fp:ap}) with Eq.~(\ref{eq:La:approx}), and performing integrations by part with vanishing boundary terms. Since we are interested in the steady state, we set the time derivative on the left hand side to zero. Note that we have a choice for the rates $\kap^\pm$ as long as they obey the condition Eq.~(\ref{eq:db:ap}). Here, we assume that the rate $\kap^+$ is a constant. After expanding $\mean{\dot n}\approx\kap^+\beta\mean{\Delta\mu-\hat f}$ we solve for $\mean{\hat f}$ and finally obtain the expression
\begin{equation}
  \label{eq:ndot:lin}
  \mean{\dot n} \approx 
  \frac{\kap^+\beta\Delta\mu}{1+(\lam/\ell)^2\kap^+\tau_0/(1+\tau_0/\tx)}
\end{equation}
valid in the linear regime $\mean{\dot n}\propto\beta\Delta\mu$ of small driving affinity. In the same linear regime, for the average reservoir work $\mean{\dot w_\text{res}}=\mean{\dot n}\Delta\mu$ we obtain
\begin{equation}
  \label{eq:wres:lin}
  \mean{\dot w_\text{res}} = 
  \left(\frac{1}{\kap^+\tau_0(\lam/\ell)^2}+\frac{1}{1+\tau_0/\tx}\right)
  \frac{\mean{\hat v}^2\tau_0}{\beta\ell^2}
\end{equation}
with average speed $\mean{\hat v}=\lam\mean{\dot n}$. For comparison, the work rate in the constant-flux ensemble (\emph{i.e.}, for ABPs) reads
\begin{equation}
  \label{eq:w:cf}
  \mean{\dot w_\text{ag}} = \frac{1}{1+\tau_0/\tx}\frac{v_0^2\tau_0}{\beta\ell^2}
\end{equation}
for all speeds $v_0$. Hence, the work spent by the reservoir is always larger than forcing a constant current $\dot n$ without fluctuations.

Using a kinetic Monte Carlo scheme for the transitions $n\leftrightharpoons n+1$ in addition to integrating the discretized Langevin equations, we have solved numerically the full stochastic dynamics of the particle \emph{and} the reactions. In the following we set $\kap^+\tau_0=10^3$ and the integration time step to $\Delta t=10^{-3}\tau_0$. The result for the average solute flux is plotted in Fig.~\ref{fig:num}(a) for three values of $\lam/\ell$. Also shown is the limiting result $\mean{\dot n}_0=\kap^+(1-e^{-\beta\Delta\mu})$ in the absence of a potential, which is approached for $\lam/\ell\to0$ (since the potential difference for directed steps becomes negligible). We see that Eq.~(\ref{eq:ndot:lin}) indeed describes the linear regime for small $\Delta\mu$. The range of validity of the linear approximation increases as $\lam/\ell$ becomes larger. In Fig.~\ref{fig:num}(b) we plot the corresponding work rate, which for sufficiently large $\lam/\ell$ and smaller speeds is well approximated by the quadratic expression Eq.~(\ref{eq:wres:lin}). Increasing $\lam/\ell$ further, the work rate approaches that of the constant-flux ensemble. Hence, the work in both ensembles becomes equivalent in the limit
\begin{equation}
  \label{eq:equiv}
  \kap^+\tau_0 \gg \frac{1+\tau_0/\tx}{(\lam/\ell)^2}
\end{equation}
of large solute flux and, consequently, small fluctuations. Note that in the opposite limit corresponding to a vanishing external potential the work Eq.~(\ref{eq:wres:lin}) is determined by the first term and thus the constant-flux approximation of ABPs is no longer valid. In Fig.~\ref{fig:num}(c), we show the average work changing the orientational correlation time $\tx$.

Choosing for the particle radius $a=\ell=1\,\mu$m, one obtains $\tau_0\approx10\,$s for water at room temperature. Hence, speeds on the order of $\mu$m/s are reached for driving affinities $\Delta\mu$ of a few $\kT$ consuming 100 reactant molecules per second. These speeds agree with what is observed for self-propulsion due to the demixing of a near-critical binary water-lutidine solvent~\cite{butt12,butt13}.

\subsection{Discussion}

\subsubsection{Neglecting translational noise}

As observed in computer simulations, the translational noise on the particle positions has little influence on the large-scale behavior, in particular one still observes a motility induced phase separation~\cite{yaou12}. This has motivated a modification of ABPs with
\begin{equation}
  \label{eq:aoup}
  \dot\X_k = -\mu_0\pd{U}{\X_k} + \vec u_k, \qquad
  \tx\dot{\vec u}_k = -\vec u_k + \nois_k
\end{equation}
called the active Ornstein-Uhlenbeck process (AOUP)~\cite{fodo16}. The noise now stems from the fluctuations of the orientations $\vec u_k$ ($\nois_k$ is Gaussian with noise strength $\Da>0$), which are not normalized anymore. The orientational correlations are still determined by $\tx$.

Conceptually, the limit $D_0\to0$ would imply $T\to0$ of the heat bath, which violates one of the basic assumptions we made in the beginning. There are two options to proceed: one can interpret Eq.~(\ref{eq:aoup}) as equations of motion arising from some non-equilibrium medium and construct thermodynamic notions in analogy to stochastic thermodynamics. This route has been followed in Refs.~\cite{fodo16,mand17} for the AOUP (see also Refs.~\cite{gang13,pugl17} for similar treatments). Both works map the coupled equations of motion to an underdamped model for which they calculate the path entropy following the standard approach of stochastic thermodynamics. These works arrive at different expressions and conclusions, which highlights the conceptual difficulties of this route. In particular, Ref.~\cite{fodo16} posits a continuation of the effective equilibrium regime to linear order of $\tx$ with vanishing path entropy production at variance with established results for the linear response regime. Moreover, even for a harmonic potential the authors predict a vanishing entropy production. Ref.~\cite{mand17} posits that an additional term besides the dissipated heat is required to restore the second law. Such a modification of the second law is not plausible for the physical mechanism underlying the directed motion and, as shown here, not necessary.

The arguably more transparent route is, for the same \emph{physical} system, to interpret these equations as effective equations of motion neglecting the translational noise. The influence of the heat bath now only enters through the dynamics of $\vec u_k$, which, for colloidal particles, we still identify with the local solvent flow. The expressions for work and heat then remain unchanged, in particular Eq.~(\ref{eq:w:abp}) is the work spent by the solvent on the particles. We calculate again the average work from Eq.~(\ref{eq:aoup}) for a single particle moving in the harmonic potential $U=\frac{1}{2}k\X^2$. For the correlations, we now obtain $\mean{\vec u\cdot\X}=\Da/(1+\tx/\tau_0)$. Choosing $\Da=v_0^2\tx$, we recover exactly the same work rate Eq.~(\ref{eq:w:cf}) as for the constant-flux ensemble of active Brownian particles.

If, instead, we control the noise strength $\Da$ and orientational correlation time $\tx$ independently as suggested in Ref.~\cite{fodo16}, we obtain
\begin{equation}
  \label{eq:w:aoup}
  \mean{\dot w_\text{ag}} = \frac{1}{1+\tx/\tau_0}\frac{\Da}{\beta\ell^2}.
\end{equation}
In Ref.~\cite{fodo16} it has been shown that in the limit $\tx\to0$ the stationary distribution $\Psi\propto e^{-\beta_\text{eff}U}$ approaches a Boltzmann distribution at an effective temperature $\kT_\text{eff}=\Da/\mu_0$. However, Eq.~(\ref{eq:w:aoup}) shows that the work and thus the dissipation do not vanish in this limit. The behavior is thus fundamentally different from active colloidal particles [cf. Fig.~\ref{fig:num}(c)], which for $\tx\to0$ reach thermal equilibrium with vanishing dissipation.

\subsubsection{Excess work}

In Ref.~\cite{spec16}, we have explored the idea that dissipation of ABPs can be modeled as an effective non-conservative force $\vec f_k=-(v_0/\mu_0)\vec e_k$. While here we have shown that the dissipation has to be modeled as the flow term Eq.~(\ref{eq:u:abp}) due to the underlying coupling to chemical reactions, the expression for the excess work (perturbing the non-equilibrium steady state) remains the same in both approaches. To this end, we insert the Langevin equations into the work
\begin{equation}
  \dot w_\text{ag} = v_0\vec e_k\cdot\pd{U}{\X_k} 
  = \vec f_k\cdot[\dot\X_k-v_0\vec e_k-\nois_k].
\end{equation}
Perturbing the particle positions $\{\X_k\}$ in the target space thus requires the excess work
\begin{equation}
  \delta w_\text{ex} = \left[\pd{H}{X_i} 
    + \vec f_k\cdot\pd{\X_k}{X_i}\right]\delta X_i
\end{equation}
with an additional term due to the work required to keep the system in the non-equilibrium steady state. Hence, all conclusions of Ref.~\cite{spec16} regarding the pressure and interfacial tension of ABPs remain valid for the identification of work and heat in the constant-flux ensemble proposed here.


\section{Conclusions and outlook}

The accurate numerical sampling of non-equilibrium steady states is a current major challenge, in particular to understand driven soft and biological materials. Here we have presented a systematic and thermodynamically consistent route to the governing equations of motion for isothermal systems that can be driven in two ways: (i)~through an external agent changing parameters with constant rate (constant-flux ensemble) or (ii)~through an ideal reservoir (constant-affinity ensemble). These two situations extend the notion of ensembles in equilibrium statistical mechanics in which either the extensive quantity is conserved or its conjugate intensive variable is fixed. In analogy with two seminal theorems in electric circuits, the two non-equilibrium ensembles sometimes go by the names of Norton and Th\'{e}venin ensemble. While numerical schemes to constrain currents, \emph{e.g.} through Gaussian cost functionals~\cite{morriss}, have been developed, our approach is based on stochastic thermodynamics and ensures that the dissipated heat $\dot q$ equals the entropy produced in the equilibrium environment, $\mathcal S=\beta q$. By construction, this equality on the level of single trajectories entails the fluctuation theorem and the (unmodified) second law. For zero and small driving, our formalism reduces to thermodynamic equilibrium ensembles and established results in the linear response regime, respectively. Moreover, we have shown that the resulting equations of motion can be decomposed into a reference system obeying detailed balance and a geometric deformation of particle positions. Such mappings between reference and target system are known from continuum mechanics and the extended ensemble approach (cf. Andersen's barostat~\cite{ande80}) but, in contrast, here the target system is steadily driven characterized by a non-vanishing average dissipation rate.

One important consequence is that the non-potential term breaking detailed balance has the nature of a ``flow'' term changing its sign with respect to time reversal (whereas non-conservative forces are invariant). This addresses the problem whether to model the driving term as a flow or force, which is not obvious from the equations of motions alone but has to be decided on physical grounds. It emphasizes that the physical cause of the driving cannot be neglected and that the reverse approach, inferring a thermodynamic description from the equations of motion of the colloidal particles alone, might yield ambiguous results.

We have developed the formalism for sheared colloidal suspensions and molecular motors, two well-studied paradigms of stochastic thermodynamics, and exemplified its usefulness applying it to the rapidly evolving field of active colloidal particles. Here the deformation is a directed translation of particles in response to each conversion of a molecular solute driven by a non-zero chemical potential difference $\Delta\mu$. A central result of our analysis is that the well-studied model of interacting active Brownian particles can be understood as the constant-flux realization of Janus particles being explicitly driven by chemical events. The corresponding work rate becomes $\dot n_\circ \Delta\mu\approx v_0\vec e_k\cdot\pd{U}{\x_k}$ in the limits of large solute fluxes and small translation distance $\lam$, both of which are fulfilled for micrometer-sized colloidal particles.

In this first step, we have neglected a spatial dependence of the concentration of molecular solutes driving the propulsion, assuming a ``pervading'' reservoir of reactant molecules. In more realistic situations, however, these molecules might only be exchanged at the system's boundary. Consumption of molecules on the particle surfaces then induces depletion and long-range concentration profiles, giving rise to phoretic interactions~\cite{yan16a,huang17}. Moreover, we have treated the solvent as a structureless ideal medium, whereas in a real fluid the solvated colloidal particles will induce correlations. Both effects could be included in the theory presented here on the level of a Gaussian field theory~\cite{spec13}. On the practical side, we have derived a simple modification of ABPs [Eq.~(\ref{eq:abp:v2})] for the constant affinity ensemble with the difference of chemical potential held fixed. The consequences for the collective dynamics and the motility induced phase transition will be explored elsewhere.

To study the collective behavior of active matter, typically coarse-grained dynamic equations are employed~\cite{nard17}. To ensure consistency with the microscopic heat dissipation, novel algorithms to systematically construct such coarser models from the microscopic equations of motion are needed. Progress in this direction has been made recently through a cycle-based approach~\cite{knoc15}. Finally, the concept of reservoirs naturally introduces intensive variables out of equilibrium~\cite{bertin06}, which might pave the way to novel numerical algorithms (``grand-canonical'' simulations with fluctuating particle number~\cite{meer17}) and help to further rationalize non-equilibrium phase coexistence~\cite{taka15a,solo17,paliwal17}.


\acknowledgments

I thank Michael E. Cates and Udo Seifert for illuminating and helpful discussions. Useful discussions during a visit of the International Centre for Theoretical Sciences (ICTS) participating in the program - Stochastic Thermodynamics, Active Matter and Driven Systems (Code: ICTS/Prog-stads2017/2017/08) are acknowledged. The DFG is acknowledged for financial support within priority program SPP 1726 (grant number SP 1382/3-2). Part of this work has been supported by the Humboldt foundation through a Feodor Lynen alumni sponsorship.


\appendix

\section{Time reversal}
\label{sec:reversal}

The stochastic action corresponding to Eq.~(\ref{eq:lang}) for the reference coordinates reads
\begin{equation}
  \label{eq:act:A}
  \mathcal A = 
  \Int{t} \frac{1}{4D_0}\sum_{k=1}^N\left(\dot\x_k+\mu_0\pd{U}{\x_k}\right)^2.
\end{equation}
Depending on stochastic calculus there are additional terms, which, however, are irrelevant for the entropy production. Denoting time reversal by $\mathcal A^\dagger$ mapping $\dot\x_k\mapsto-\dot\x_k$, the part of the action that is asymmetric under time reversal is identified with the (dimensionless) entropy production
\begin{equation}
  \label{eq:act:S}
  \mathcal S = \mathcal A^\dagger - \mathcal A 
  = -\beta\Int{t}\pd{U}{\x_k}\cdot\dot\x_k = \beta q,
\end{equation}
which equals the heat $q$ [as identified from Eq.~(\ref{eq:first})] dissipated into the heat bath at inverse temperature $\beta$. This agreement guarantees the consistency of stochastic thermodynamics since the heat appearing in the first law is the same heat determining the second law.

\section{Hydrodynamic interactions}
\label{sec:hydro}

Including hydrodynamic coupling, the Langevin equation~(\ref{eq:lang}) for the reference positions becomes
\begin{equation}
  \dot\x_k = -\mmu_{kl}\cdot\pd{U}{\x_l} + \nois_k,
\end{equation}
where the symmetric mobility matrices $\mmu_{kl}$ depend on particle separations. The same mobility matrices now determine the noise correlations
\begin{equation}
  \label{eq:hyd}
  \mean{\nois_k(t)\nois_l^T(t')} = 2\kT\mmu_{kl}\delta(t-t')
\end{equation}
so that the stochastic action reads
\begin{equation}
  \mathcal A = \frac{\beta}{4}\Int{t}
  \left(\dot\x_k+\mmu_{ki}\cdot\pd{U}{\x_i}\right)\cdot\mmu^{-1}_{kl}\cdot 
  \left(\dot\x_l+\mmu_{li}\cdot\pd{U}{\x_i}\right)
\end{equation}
with $\mmu_{ki}\cdot\mmu^{-1}_{il}=\delta_{kl}\id$. Calculating the asymmetric contribution Eq.~(\ref{eq:act:S}), we find the same result as in the absence of hydrodynamic interactions. This demonstrates that, as long as Eq.~(\ref{eq:hyd}) is fulfilled, the dissipation along a \emph{single trajectory} is not influenced by the hydrodynamic coupling, see also Ref.~\cite{spec08}.

\section{Derivation of work relation Eq.~(\ref{eq:ft:ca})}
\label{sec:ft}

For the derivation of Eq.~(\ref{eq:ft:ca}), we adopt the method considering the time evolution of a transformed joint probability of state and work~\cite{jarz97a,spec04,impa05a}. First, we recast Eq.~(\ref{eq:fp}) as $\partial_t\psi=\mathcal L_\text{ref}\psi$ defining the evolution operator $\mathcal L_\text{ref}$ with stationary solution $\psi_\text{eq}$. The work rate (in this section we drop the subscript to ease notation) reads
\begin{equation}
  \dot w = f_i\dot X_i = \Gamma\beta f_i(f_i-\hat f_i) + f_i\zeta_i
\end{equation}
inserting Eq.~(\ref{eq:Xdot}). The evolution equation for the joint probability $\phi(\z,\vec X,w;t)$ of state and accumulated work becomes
\begin{equation}
  \label{eq:phi}
  \partial_t\phi = \mathcal L_\text{ref}\phi 
  - \Gamma\beta f_i(f_i-\hat f_i)\pd{\phi}{w} 
  + \Gamma f_i\pd{^2\phi}{X_i\partial w}
\end{equation}
since the work and the exchanged quantities share the same noise. We define the transformed $\hat\phi(\z,\vec X;t)\equiv\Int{w}\phi(\z,\vec X,w;t)e^{-\beta w}$ with initial condition $\hat\phi(\z,\vec X;0)=\psi_\text{eq}(\z,\vec X)$. The evolution equation becomes
\begin{equation}
  \partial_t\hat\phi = \mathcal L_\text{ref}\hat\phi 
  + \Gamma\beta f_i\left[-\beta(f_i-\hat f_i)\hat\phi + \pd{\hat\phi}{X_i}\right]
\end{equation}
after inserting Eq.~(\ref{eq:phi}) and following integrations by parts with respect to the work $w$. The solution of this equation obeying the initial condition is the Boltzmann distribution [Eq.~(\ref{eq:boltz})] $\hat\phi=\psi_\text{eq}$ independent of $t$. We stress that the actual probability distribution $\psi(\z,\vec X;t)$ for $t>0$ is different from the Boltzmann distribution. Hence, we obtain
\begin{equation}
  \mean{e^{-\beta w_\text{res}}} = \sum_{\z,\vec X} \hat\phi(\z,\vec X) = 1
\end{equation}
for a system starting in thermal equilibrium but reaching a steady state due to affinities $\vec f$ that are not attainable in equilibrium.


%

\end{document}